# PQC: Triple Decomposition Problem Applied To *GL(d, F$_p$)* - A Secure Framework For Canonical Non-Commutative Cryptography

P. Hecht[1]

*Abstract*— Post-Quantum Cryptography (PQC) attempts to find cryptographic protocols resistant to attacks using Shor's polynomial time algorithm for numerical field problems or Grover's search algorithm. A mostly overlooked but valuable line of solutions is provided by non-commutative algebraic structures, specifically canonical protocols that rely on one-way trapdoor functions (OWTF). Here we develop an algebraic framework who could be applied to different asymmetric protocols like D-H KE (Diffie-Hellman key exchange), Public Key Encryption, Digital Signature, ZKP (zero-knowledge proof) authentication, Oblivious Transfer, Multi-Party Computing, and so on. The trapdoor one-way functions selected are (a) Triple decomposition Problem (TDP) developed by Kurt, where a known element is factored into a product of three unknown factors and (b) a new version of conjugacy search that we refer from now on as Blind Conjugacy Search Problem (BCSP). Our platform structure is the general linear group *GL(d, F$_p$)* d-square non-singular matrices of prime field values. In our case, we use p=251 (the biggest prime fitting into a byte) and d=8. We give support to the fact that this framework is cryptographically secure against classical attacks like linear algebra attacks, length-based attacks, side-channel attacks against square (or duplicate) and multiply (or sum) algorithm, high sensitivity to pseudo random deterministic generators, etc. At same time it is immune against quantum attacks (using Grover and Shor), if the size parameters are carefully selected. Also, there is no pattern emerging from the intermediate computations, and therefore resulting tokens or encrypted information is indistinguishable from same size random matrices. This feature is a key point to assign semantic security to the framework and a way to obtain IND-CCA2 security. An additional advantage is that all operations fit into byte modular arithmetic's.

## 1. INTRODUCTION

Post-Quantum Cryptography (PQC) is a trend that has an official NIST status [1,2] and which aims to be resistant to quantum computers attacks like Shor [3] and Grover [4] algorithms. NIST initiated last year a process to solicit, evaluate, and standardize one or more quantum-resistant public-key cryptographic algorithms [1]. Particularly Shor algorithm provided a quantum way to break asymmetric protocols.

Security of a canonical non-commutative protocol always relies his security on a one-way trapdoor function (OWTF) [5]. For instance, in an algebraic context, the conjugacy search problem, decomposition problem, double coset problem, triple decomposition problem, factorization search problem, commutator based or simultaneous conjugacy search problem. All are hard problems assumed to belong to AWPP time-complexity (but out of BQP) [6] which yield convenient computational security against current quantum attacks.

More information about post-quantum cryptography (PQC), non-commutative cryptography (NCC) and non-associative cryptography (NAC) could be found in published works [7,8,9,10,11,12,13].

## 2. THE PLATFORM GROUP: *GL(d, F$_p$)*

The algebraic platform uses the general linear group, non-singular modular matrices of d-dimension.

Let *p* be a prime, *d* any integer >1, $q=p^d$ and *F$_p$[x]* the polynomial extension of the prime field *F$_p$*. The number of square matrices of order *d* and values in *F$_p$* is $p^{d^2}$, and of those $p^{d^2-d}$ are nilpotent [14]. The number of elements in the general linear group of *d*-order non-singular square matrices is:

$$|GL(d, F_p)| = \prod_{i=0}^{d-1}(p^d - p^i) \quad (1)$$

A non-singular matrix or *d*-order whose monic characteristic polynomial is irreducible in *F$_p$*, generates a cyclic (thus commutative) subgroup $P_d$ of $M_d = GL(d, F_p)$. Each *d*-degree irreducible polynomial *f(x)* in *F$_p$[x]* field has a square companion matrix of *d*-order who acts as a generator of the multiplicative cyclic subgroup $P_d$, and each member of this subgroup corresponds to a unique monic characteristic polynomial of at most *d-1* degree [14]. The $N_{tot}$ number of non-trivial (null or unitary) monic *d*-degree *f(x)* over *F$_{251}$* field is:

$$N_{tot} = p^d - 2 \quad (2)$$

Using Möbius $\mu$ function, the $N_p(d)$ number of monic irreducible *d*-degree polynomials over *F$_p$[x]* field is:

$$N_p(d) = \frac{1}{d}\sum_{r|d} \mu(d)p^{d/r} = \frac{p^d - 2}{d} = \frac{N_{tot}}{d} \quad (3)$$

To generate a random *d*-order monic irreducible polynomial over *F$_p$[x]*, we use the probabilistic Algorithm 4.70 [15] whose complexity is $O(m^3 (lg\ m)(lg\ p))$ and requires approximately *d*-trials. Once found, it is translated into the companion matrix. It is of interest to find its order, because that would be the number of elements of the commutative subgroup $P_d$ of the $M_d$ matrix group. Whatever this value is, it must be a divisor of the multiplicative subfield order ($= p^d - 1$) and if it were maximal, the irreducible polynomial would be a primitive one. To calculate polynomial orders, a modified version of Algorithm 4.77 [15] can be used.

---

[1] Pedro Hecht: Maestría en Seguridad Informática, FCE-FCEyN-FI (Universidad de Bs Aires) phecht@dc.uba.ar

Using an irreducible polynomial in an extension field is a method of generating a $P_d$ commutative subgroup of the non-singular modular square matrices, but there exists another way to achieve the same goal. For matrices, the necessary and sufficient condition for two symmetric (diagonalizable) matrices to commute, is that they share the same orthonormal basis, that means the same eigenvectors $P$ matrix [16]. If we start from two different diagonal matrices with strictly positive eigenvalues $D_1$, $D_2$; then the transformed $A = P^{-1} D_1 P$ and $B = P^{-1} D_2 P$ commute. The later approach is computational much faster than the first one, so it is followed in our framework.

For the $p=251$, $d=8$ example, the involved cardinals are depicted at Table I.

TABLE I
PARTICULAR SETS SIZES

| Set | Cardinal |
|---|---|
| (mod 251) 8-dim square matrices | $3.794182134705598 \times 10^{153}$ |
| General Linear Group $GL(8, F_{251})$ | $3.779005647067214 \times 10^{153}$ |
| Singular (mod 251) 8-dim square matrices | $1.517648763838442 \times 10^{151}$ |

To generate a $GL(8, F_{251})$ matrix, a random (mod 251) 8-dim matrix is obtained and rejected if its determinant were null, which occurs in about 4% of the cases, a new random matrix is found. The same non-singularity control is applied before attempting any further matrix inversion in the protocol, as shown in the APPENDIX I source code.

3. OWTF: TRIPLE DECOMPOSITION PROBLEM (TDP)

As usual, $[a,b]=a^{-1}b^{-1}ab$, is the multiplicative commutator of $a,b$. The TDP protocol, is stated as follows.

**TDP protocol:**
Given a G group (or monoid) and two sets of five subsets of G, say $A=\{A_1, A_2, A_3, X_1, X_2\}$ and $B=\{B_1, B_2, B_3, Y_1, Y_2\}$, satisfying the following conditions:

**(Invertibility conditions)** The elements of $X_1, X_2, Y_1, Y_2$ are invertible.

**(Commutativity conditions)** $[A_2,Y_1]= [A_3,Y_2]= [B_1,X_1]= [B_1,X_1]= [B_2,X_2]= I$ (identity)

Alice and Bob agree on who will use which sets of subsets, say Alice uses A and Bob uses B. Then the exchange between both goes as follows:

(1) Alice chooses $a_1 \in A_1$, $a_2 \in A_2$, $a_3 \in A_3$, $x_1 \in X_1$, $x_2 \in X_2$, and computes $u=a_1 x_1$, $v=x_1^{-1} a_2 x_2$, and $w=x_2^{-1} a_3$. Her private key is $(a_1, a_2, a_3)$.

(2) Bob chooses $b_1 \in B_1$, $b_2 \in B_2$, $b_3 \in B_3$, $y_1 \in Y_1$, $y_2 \in Y_2$, and computes $p=b_1 y_1$, $q=y_1^{-1} b_2 y_2$, and $r=y_2^{-1} b_3$. His private key is $(b_1, b_2, b_3)$.

(3) Alice sends the public key $(u, v, w)$ to Bob.

(4) Bob sends the public key $(p, q, r)$ to Alice.

(5) Alice computes
$K_A = a_1 p a_2 q a_3 r = a_1(b_1 y_1) a_2(y_1^{-1} b_2 y_2) a_3(y_2^{-1} b_3) = a_1 b_1 a_2 b_2 a_3 b_3$

(6) Bob computes
$K_B = u b_1 v b_2 w b_3 = (a_1 x_1) b_1 (x_1^{-1} a_2 x_2) b_2 (x_2^{-1} a_3) b_3 = a_1 b_1 a_2 b_2 a_3 b_3$

**Computational TDP:**
**Given the TDP protocol, compute any component of Alice's (or Bob's) private key (k1, k2, k3) given her (or his) public key (x, y, z).**

The Yesem Kurt TDP protocol is fully described in [17]. We use the Protocol I there described as we found it better fitted to our platform.

4. OWTF: BLIND CONJUGACY SEARCH PROBLEM (BCSP)

The computational Conjugacy Search Problem (CSP) is:

**Computational CSP:**
**Given G, a non-abelian group with solvable word problem and given two elements $a, b \in G$, find at least one element $x \in G$ such that $a = x^{-1} b \; x$.**

From this problem we derive a stronger one, the computational Blind Conjugacy Search Problem (BCSP) as follows:

**Computational BCSP:**
**Given G, a non-abelian group with solvable word problem and given any element $a \in G$, and an unknown element $b \in G$, find at least one element $x \in G$ such that $a = x^{-1} b \; x$.**

The additional challenge here is to find a conjugator ignoring the conjugated element. Any time two entities share a common key element of a group, if that key is used as a conjugator, it can be used to hide (or encrypt) any other conjugated element. In our TDP based framework, we use the shared key as conjugator for any matrix encoded plain text.

5. STEP-BY-STEP DESCRIPTION

We present a combined D-H and BCSP cipher protocol, the only difference between both lies at the last step added.

In our version, we work with two entities (Alice and Bob), but this could be easily generalized for any number of participants. All arithmetic operations should be assumed belonging to field $F_{251}$.

TABLE II
SYMBOLS AND DEFINITIONS.

| |
|---|
| $M_8 \equiv GL(8, F_{251})$ – selected parameters are $d=8$ and prime=251 |
| $P_8 \subset M_8$ – any commutative subgroup |
| $\subset$ – strictly included into |
| $\in$ – belongs to |
| $\in_R$ – uniform distribution, randomly selected element in |
| $\forall\neq$ - strictly positive elements in the list. |
| $d_A, d_B, ...$ – diagonal matrices of eigenvalues with $\forall\neq$ property |
| $(\lambda_1 ... \lambda_8)$ – eigenvalues set, each one mentioned independent from others |
| $[a,b]$ – commutator ($=a^{-1}b^{-1}ab$) |
| $I$ – Identity matrix order 8 |
| $Sel$ – selects or reserves for her/him with agreement of the other party. |
| $\Rightarrow$ send publicly to the other entity |
| validation – greyed field means a consistency proof |

TABLE III
PUBLIC SETUP STEPS

| | |
|---|---|
| Public parameters (any entity defines) | Subgroups $\{A_1, A_2, A_3, X_1, X_2\} \subset M_8$<br>Subgroups $\{B_1, B_2, B_3, Y_1, Y_2\} \subset M_8$<br>Eigenvectors $P$ assigned to $A_2$, $Y_1$ ; $[A_2, Y_1]=I$<br>Eigenvectors $Q$ assigned to $A_3$, $Y_2$ ; $[A_3, Y_2]=I$<br>Eigenvectors $R$ assigned to $B_1$, $X_1$ ; $[B_1, X_1]=I$<br>Eigenvectors $S$ assigned to $B_2$, $X_2$ ; $[B_2, X_2]=I$<br>Eigenvectors matrices $\{P, Q, R, S\} \in_R M_8 \Rightarrow$ |

TABLE IV
PRIVATE PROCEDURES

| | ALICE | BOB |
|---|---|---|
| Generating private elements | $Sel\ \{A_1, A_2, A_3, X_1, X_2\}$<br>$\forall\neq \lambda_1 ... \lambda_8 \in_R \mathbb{Z}_{251}^*$<br>$d_{A2} = (\lambda_1 ... \lambda_8)$<br>$d_{A3} = (\lambda_1 ... \lambda_8)$<br>$d_{X1} = (\lambda_1 ... \lambda_8)$<br>$d_{X2} = (\lambda_1 ... \lambda_8)$<br>$x_1 = R^{-1} d_{x1} R \in P_8$<br>$x_2 = S^{-1} d_{x2} S \in P_8$ | $Sel\ \{B_1, B_2, B_3, Y_1, Y_2\}$<br>$\forall\neq \lambda_1 ... \lambda_8 \in_R \mathbb{Z}_{251}^*$<br>$d_{B1} = (\lambda_1 ... \lambda_8)$<br>$d_{B2} = (\lambda_1 ... \lambda_8)$<br>$d_{Y1} = (\lambda_1 ... \lambda_8)$<br>$d_{Y2} = (\lambda_1 ... \lambda_8)$<br>$y_1 = P^{-1} d_{y1} P \in P_8$<br>$y_2 = Q^{-1} d_{y2} Q \in P_8$ |

TABLE V
PRIVATE KEYS

| | ALICE | BOB |
|---|---|---|
| Private keys | $a_1 \in_R M_8$<br>$a_2 = P^{-1} d_{A2} P$<br>$a_3 = Q^{-1} d_{A3} Q$ | $b_1 = R^{-1} d_{B1} R$<br>$b_2 = S^{-1} d_{B2} S$<br>$b_3 \in_R M_8$ |

TABLE VI
PUBLIC KEYS

| | ALICE | BOB |
|---|---|---|
| Public keys (tokens) | $u = a_1 x_1$<br>$v = x_1^{-1} a_2 x_2$<br>$w = x_2^{-1} a_3$<br>$(u, v, w) \Rightarrow$ | $p = b_1 y_1$<br>$q = y_1^{-1} b_2 y_2$<br>$r = y_2^{-1} b_3$<br>$(p, q, r) \Rightarrow$ |

TABLE VII
COMMON D-H SESSION KEY

| | ALICE | BOB |
|---|---|---|
| Private session key ($K$) obtained | $K_{ALICE} =$<br>$= a_1\ p\ a_2\ q\ a_3\ r$ | $K_{BOB} =$<br>$= u\ b_1\ v\ b_2\ w\ b_3$ |
| | $K_{ALICE} = a_1\ p\ a_2\ q\ a_3\ r =$<br>$= a_1 (b_1 y_1) a_2 (y_1^{-1} b_2 y_2) a_3 (y_2^{-1} b_3) =$<br>$= a_1\ b_1\ a_2\ b_2\ a_3\ b_3)$<br>$K_{BOB} =\ u\ b_1\ v\ b_2\ w\ b_3$<br>$= (a_1 x_1) b_1 (x_1^{-1} a_2 x_2)\ b_2 (x_2^{-1} a_3) b_3 =$<br>$= a_1\ b_1\ a_2\ b_2\ a_3\ b_3$<br>$K_{ALICE} = K_{BOB} \equiv\ K$ | |

TABLE VIII
ADDITIONAL ENCRYPTION STEP (BCSP PROTECTED)

| | ALICE | BOB |
|---|---|---|
| BOB ciphers a message for ALICE | | $msg \in M_8$<br>$cif =$<br>$= K_{BOB}^{-1} msg\ K_{BOB}$<br>$cif \Rightarrow$ |

| | ALICE | BOB |
|---|---|---|
| ALICE recovers the message | $msg =$<br>$= K_{ALICE}\ cif\ K_{ALICE}^{-1}$ | |
| | $msg = K_{ALICE}\ cif\ K_{ALICE}^{-1} =$<br>$= K_{ALICE}(K_{BOB}^{-1} msg\ K_{BOB})K_{ALICE}^{-1} = msg$ | |

It is easy to apply the same framework to other asymmetric protocols. For example, defining power-sets of matrices, a straightforward ElGamal solution is at hand. Also, extending the GL() to a polynomial ring, a Maze et al. protocol could be implemented [8]. Changing of purpose, a ZKP authentication protocol, a Baumslag et al. KEM (key encapsulation mechanism) or a Digital Signature is almost trivial to design [9]. When powers of matrices are used new kind of weakness could appear, the multiplicative order of the elements should be sufficiently high to foil brute-force attacks. One could use companion matrices of primitive polynomials as generators of high order subgroups. Comparing different canonical approaches, the BCSP application of this framework offer a reasonable compromise solution between cryptographic security and fast computation.

At APPENDIX I we present a standard Mathematica source code and the running results (on a Core i5 PC/2.20GHz). No attempt was made to optimize that code and computations are only included as a proof of concept. All symbols included there, follow here explained conventions.

6. FRAMEWORK SECURITY

Some required properties that the algebraic platform ($G$) should have in order to have an efficient and secure framework [8,17], could be resumed here:

I. *G* should be non-abelian and of exponential growth as function of the length of its elements. In our framework, *G* is the general linear group and the *A, B* subgroups cardinality growth exponentially as the d-dimension parameter increases linearly.
II. The word problem (determining if an element is or not an identity) should be efficiently solved. That achieves trivially the matrix framework.
III. Elements of *G* should be efficiently represented on a computer. This is also trivial in our case.
IV. Multiplication and inversion of elements should be computationally easy within the representation. Again, the proposed framework complies it.
V. If centralizers of generator sets are used, it should be hard to solve a decomposition problem presented in [17]. In our case, we do not require them.

As shown, our framework fits well into the required settings.

A way to attack the present protocol would be to find a probabilistic polynomial time algorithm to solve the algebraic triple decomposition problem and the blind conjugacy search problem.

To guarantee that TDP truly rely on non-linear triple decomposition, the framework should avoid some pitfalls. One way to attack TDP is to find a pseudo-key that works as a private one of any entity [17], which requires solving:

$$a_1 x_1 = u \quad (4)$$
$$x_1^{-1} a_2 x_2 = v \quad (5)$$
$$x_2^{-1} a_3 = w \quad (6)$$

Specifically, equation (5) is quadratic in terms of three unknowns and this foils linear algebraic attacks. There should be paid attention to some cases to be avoided. We resume them here.

(a) If $[X_1,Y_1]=I$, $[X_2,Y_1]=I$ and $[X_2,Y_2]=I$, then the shared key could be computed from the public key.
(b) If $[A_2,B_1]=I$, $[A_3,B_2]=I$ and $[A_3,B_1]=I$, then the shared key could be computed from the public key.
(c) If ($[A_2,B_1]=I$ and $[X_2,B_1]=I$) or ($[A_3,B_2]=I$ and $[A_3,Y_1]=I$), then the security of the system relies on the difficulty of decomposing an element into two unknowns.
(d) To assure that eq. (5) cannot be linearly reduced, eq. (6) should have many solutions.
(e) In case of the use of generators to define the subgroups in the setting, they should be short to foil a length-based attack.

More details of these considerations could be found at Kurt's paper [17] and Myasnikov et al. book [8]. In our framework:

(a) Large cardinality of involved sets could be obtained selecting appropriate dimensional parameters.
(b) Commutativity restrictions are accurately followed as each eigenvector matrix are independent of others.
(c) Equation (6) has a very big number of solutions (see eq. (7)).
(d) No generators are used to define the subgroups sets.

Supposing no other weakness at hand, full cracking a private key depends on four *d*-dimensional diagonal eigenvalues matrices, so a brute force search of the commutative $P_8$ subgroups of $M_8$ involves the cardinal

$$|P_8|^4 = 249^{32} = 4.77 \times 10^{76} \approx 2^{255} \quad (7)$$

For a real-life application, we suggest to use at least $P_8$ or perhaps expanding the commutative subgroup to $P_{16}$, who implies a *510*-bit level classical security. The corresponding quantum security level is the respective square roots of the key space cardinals [7]. The obvious drawback is the corresponding increase in space, each matrix in $\{P_8, P_{16}\}$ occupies respectively *{512, 2048}* bits. A corresponding computational session time should be expected.

Against quantum-based attacks, the dimensional increase foils Grover like attacks and no multiplicative (or additive) cyclic order finding adaptation of Shor's algorithm is known and accessible against the present protocol.

As a final consequence, we infer that our proposed framework could be gotten immune against brute-force attacks, linear representation attacks, length-based attacks and currently known quantum attacks.

7. SEMANTIC SECURITY

The key point to assure semantic security is the indistinguishability of encrypted information from random one of same length [5,18,19,20].

The presented framework is easily translatable to other asymmetric protocols. For this reason, the following security analysis is not limited to the present example and can be extended in new contexts. To proceed with this analysis, two conjectures are exposed.

***DEFINITION (D1): interactive challenge-response game by a verifier against an active adversary.***

In this three-phase protocol, two entities, an adversary and a verifier (or challenger) are involved. The verifier has a secret key that he tries to hide from the adversary and allows the adversary to pose questions to him that answers truthfully like an Oracle. In a first phase, the adversary can raise all the questions that he wants to try to obtain information about the secret key. In a second phase, the verifier presents the secret key (*k*) to the adversary next to another of equal length and format (*\* k*) randomly generated. Even during the second phase, the adversary may continue to consult the verifier, except for questions linked to the disclosure of the secret itself. In the third phase, the adversary has a polynomial time stochastic algorithm and must distinguish whether the secret is *k* or *\* k*, with probability negligibly greater than ½. If the opponent achieves the distinction with that probability, he wins the game and loses it in the opposite case.

***CONJECTURE (C1): Indistinguishability of product transformed random matrices.***

The elements of $GL(d, F_p)$ are uniformly random integers of prime modulus and d-dimension. It is a known fact that sum or multiplication between random field integers, does not introduce statistical bias into results [21]. Therefore, linear transformed matrices are statistically distributed as any random generated ones. The consequence is that in an interactive challenge-response protocol (Definition D1), an adversary does not achieve the distinction raised with the required probability.

***CONJECTURE (C2): the present framework adheres to semantic security under IND-CCA2.***

The TDP one-way trapdoor function with which the private key is protected is, as previous exposed, not weakened by attacks of probabilistically polynomial time, whether classical or quantum that are in the public domain until today, forcing the potential attacker to perform a systematic exploration of the private key space (the diagonal matrices). Under this assumption and considering the indistinguishability of randomly generated matrices and enciphered ones, it is reasonable to assign to the framework a security mark equivalent to IND-CCA2 (semantic security under IND-CCA2) [19,20].

## 8. Previous Work

Some previous work over the same issue was conducted by the author along last year's [22,23,24]. Recently we contributed to NIST standardization Program with a Maze et al. protocol where the selected platform was based in octonions algebra [25]. Unfortunately, it was a bad choice because an algebraic attack of Bernstein [26] exposed the octonion's weakness. This shows that even a viable OWTF does not cover against a weak platform. Statements like the exposed in 6. Framework Security (I. to V.) should be faithfully pursued to achieve an efficient asymmetric cryptographic system.

## 9. Conclusions

We developed a non-arbitrated algebraic post-quantum framework of canonical non-commutative cryptography, which could easily be adapted to any other asymmetric purpose.

Security and computational efficiency were the main concerns at developing time. Therefore, TDP was selected as an appropriate OWTF for D-H Key Exchange between two entities. Once obtained, the common key is used to cipher messages between both using a simple transformation protected by BCSP, a blind conjugacy problem OWTF.

Classical and quantum attacks are discussed, and the framework seems to be resilient to both. Also, evidence to assign semantic security (IND-CCA2) to the framework is provided. Nevertheless, it must be kept attention to any appearing cryptanalytic tools against non-commutative platforms [27].

No big number library is required here. This feature would enable the use at low computational resources environments like smartcards or cryptographic keys.

## 10. References

[1] NIST, "Post-Quantum Cryptography Standardization Program", https://csrc.nist.gov/Projects/Post-Quantum-Cryptography
[2] L. Chen et al, NISTIR 8105, "Report on Post-Quantum Cryptography",NIST,2006. http://nvlpubs.nist.gov/nistpubs/ir/2016/NIST.IR.8105.pdf (consulted February10, 2017)
[3] P. Shor, "Polynomial-time algorithms for prime factorization and discrete logarithms on a quantum computer", SIAM J. Comput., no. 5, pp. 1484-1509, 1997.
[4] L. K. Grover, "A fast quantum mechanical algorithm for database search", In Proc. 28th Ann. ACM Symp. on Theory of Computing (ed. Miller, G. L.) 212–219, ACM, 1996.
[5] J. Katz, Y. Lindell, "Introduction to Modern Cryptography", Chapman & Hall/CRC, 2008.
[6] L. Fortnow, J. Rogers, "Complexity Limitations on Quantum Computation", arXiv:cs/9811023 [cs.CC], 1998
[7] D. J. Bernstein, J. Buchmann, E. Dahmen, "Post-Quantum Cryptography", Springer Verlag, ISBN: 978-3-540-88701-0 , 2009.
[8] A. Myasnikov, V. Shpilrain, A. Ushakov, "Non-commutative Cryptography and Complexity of Group-theoretic Problems", Mathematical Surveys and Monographs, AMS Volume 177, 2011
[9] L. Gerritzen et al (Editors), "Algebraic Methods in Cryptography", Contemporary Mathematics, AMS, Vol. 418, 2006
[10] B. Tsaban, "Polynomial time solutions of computational problems in non-commutative algebraic crypto", 2012. http://arxiv.org/abs/1210.8114v2, (consulted February10, 2017)
[11] M. I. González Vasco, R. Steinwandt, "Group Theoretic Cryptography" , CRC Press, 2015
[12] A. Kalka, "Non-associative public-key cryptography", 2012. arXiv:1210.8270 [cs.CR] (consulted February10, 2017)
[13] V.A. Shcherbacov, "Quasigroups in cryptology", Computer Science Journal of Moldova, 17:2, 50, 2009.
[14] R. Lidl and H. Niederreiter, "Finite Fields", Cambridge University Press, Cambridge, 1997.
[15] A. Menezes, P. van Oorschot and S.Vanstone, "Handbook of Applied Cryptography", CRC Press, 1997.
[16] T. Beth et al,, " Encyclopedia of Mathematics and its Applications", Vol 69: "Design Theory", 2 nd . Ed, Cambridge University Press, 1999
[17] Y. Kurt, "A new key exchange primitive based on the triple decomposition problem", preprint, Available at http://eprint.iacr.org/2006/378
[18] M. Bellare, A. Dessai, D. Pointcheval, P. Rogaway, "Relations Among Notions of Security for Public-Key Encryption Schemes", Advances in Cryptology (CRYPTO '98), Lecture Notes in Computer Science Vol. 1462, H. Krawczyk ed., Springer-Verlag, 1998.
[19] E. Kiltz, J. Malone-Lee, "A General Construction of IND-CCA2 Secure Public Key Encryption", available at homepage.ruhr-uni-bochum.de/Eike.Kiltz/papers/general_cca2.ps
[20] S. Goldwasser, S. Micali, "Probabilistic Encryption", Journal of Computer and System Sciences, 28: 270-299, 1984.
[21] M. D. Springer, "The algebra of random variables", Wiley series in probability and mathematical statistics, 1979
[22] P. Hecht, "Post-Quantum Cryptography(PQC): Generalized ElGamal Cipher over GF(251^8)", ArXiv Cryptography and Security (cs.CR) http://arxiv.org/abs/1702.03587 6pp (2017) & Journal of Theoretical and Applied Informatics (TAAI), 28:4, pp 1-14 (2016), http://dx.doi.org/10.20904/284001
[23] P. Hecht, "Post-Quantum Cryptography: A Zero-Knowledge Authentication Protocol", ArXiv Cryptography and Security (cs.CR) https://arxiv.org/abs/1703.08630, 3pp (2017)
[24] P. Hecht, "Post-Quantum Cryptography: S381 Cyclic Subgroup of High Order", ArXiv Cryptography and Security (cs.CR) http://arxiv.org/abs/1704.07238 (preprint) & International Journal of Advanced Engineering Research and Science (IJAERS) 4:6, pp 78-86 (2017), https://dx.doi.org/10.22161/ijaers.4.6.10
[25] P. Hecht, J. Kamlofsky, "HK17: Post Quantum Key Exchange Protocol Based on Hypercomplex Numbers", https://csrc.nist.gov/CSRC/media/Projects/Post-Quantum-Cryptography/documents/round-1/submissions/HK17.zip
[26] Bernstein D., Lange T., "HK17 Official Comment", https://csrc.nist.gov/CSRC/media/Projects/Post-Quantum-Cryptography/documents/round-1/official-comments/HK17-official-comment.pdf, 2017
[27] V. Roman'kov,"Cryptanalysis of a combinatorial public key crypto-system", De Gruyter, Groups Complex. Cryptology. 2017.

# APPENDIX I: Mathematica source code and a step-by-step session

## NEWLY DEFINED FUNCTIONS

```
RandomMatrix[dim_,prime_] := Module[{x},
    Label[1];
    x=Table[Random[Integer,{0,prime-
1}],{i,1,dim},{j,1,dim}];
    If[Det[x,Modulus→prime]==0,Goto[1],Break];x];
RandomDiagonalMatrix[dim_, prime_] :=
    DiagonalMatrix[Table[Random[Integer,{1,prime-
1}],{i,1,dim}]];
InverseMatrix[M_,prime_]:=Inverse[M,
Modulus→prime];
ConjugateMatrix[M_,conj_,prime_]:=
    Mod[InverseMatrix[conj,prime].M.conj,prime];
M[X_]:=MatrixForm[X];
```

## COMBINED D-H AND CIPHER SESSION

```
RandomSeed;
TimesToRepeat=1000;
dim=8;
prime=251;
trials=0;
{T0=TimeUsed[],Do[
    {
     Module[{},
       Label[begin];
       trials +=1;
       P=RandomMatrix[dim,prime];
       Q=RandomMatrix[dim,prime];
       R=RandomMatrix[dim,prime];
       S=RandomMatrix[dim,prime];
       d_{A2}=RandomDiagonalMatrix[dim,prime];
       d_{A3}=RandomDiagonalMatrix[dim,prime];
       d_{X1}=RandomDiagonalMatrix[dim,prime];
       d_{X2}=RandomDiagonalMatrix[dim,prime];
       x_1=ConjugateMatrix[d_{X1},R,prime];
       x_2=ConjugateMatrix[d_{X2},S,prime];
       If[Det[x_1.x_2,
Modulus→prime]==0,Goto[begin],nil;];
       a_1=RandomMatrix[dim,prime];
       a_2=ConjugateMatrix[d_{A2},P,prime];
       a_3=ConjugateMatrix[d_{A3},Q,prime];
       If[Det[a_1.a_2.a_3,
Modulus→prime]==0,Goto[begin],nil;];
       u=Mod[a_1.x_1 ,prime];
       v=Mod[InverseMatrix[x_1,prime].a_2.x_2,prime];
       w=Mod[InverseMatrix[x_2,prime].a_3,prime];
       d_{B1}=RandomDiagonalMatrix[dim,prime];
       d_{B2}=RandomDiagonalMatrix[dim,prime];
       d_{Y1}=RandomDiagonalMatrix[dim,prime];
       d_{Y2}=RandomDiagonalMatrix[dim,prime];
       y_1=ConjugateMatrix[d_{Y1},P,prime];
       y_2=ConjugateMatrix[d_{Y2},Q,prime];
       If[Det[y_1.y_2,
Modulus→prime]==0,Goto[begin],nil;];
       b_1=ConjugateMatrix[d_{B1},R,prime];
       b_2=ConjugateMatrix[d_{B2},S,prime];
       b_3=RandomMatrix[dim,prime];
       If[Det[b_1.b_2.b_3,
Modulus→prime]==0,Goto[begin],nil;];
       p=Mod[b_1.y_1 ,prime];
       q=Mod[InverseMatrix[y_1,prime].b_2.y_2,prime];
       r=Mod[InverseMatrix[y_2,prime].b_3,prime];
       Kalice= Mod[a_1.p.a_2.q.a_3.r,prime];
       Kbob= Mod[u.b_1.v.b_2.w.b_3,prime];
(* From here on the encryption routine *)
       msg=RandomMatrix[dim,prime];
       cif=ConjugateMatrix[msg,Kbob,prime];
rec=ConjugateMatrix[cif,InverseMatrix[Kalice,prime]
      ,prime];
     ];
   },{TimesToRepeat}],T1=TimeUsed[]};
```

## LAST SESSION PRINTING SOURCE

```
Print[" -------------------------------------------
-------"];
Print[" Matrix dimension (dim)   = ",dim];
Print[" Prime modulus (prime)    = ",prime];
Print[" Number of Sessions       = ",TimesToRepeat];
Print[" Mean session time (sec)  = ",TimesPerCycle];
Print[" Sessions singularity corrections needed = ",
    (trials/TimesToRepeat).100," percent"];
Print[" "];
Print[" Ongoing, the last computed session is
presented."];
Print[" "];
Print[" Random public eigenvectors -----------------
--------"];
{Print[" P (for A2,Y1) = ",M[P]],Print[" Q (for
A3,Y2) = ",M[Q]]};
{Print[" R (for B1,X1) = ",M[R]],Print[" S (for
B2,X2) = ",M[S]]};
Print[" Random ALICE private eigenvalues ----------
--------"];
{Print[" dA2 = ",M[d_{A2}]],Print[" dA3 = ",M[d_{A3}]]};
{Print[" dX1 = ",M[d_{X1}]],Print[" dX2 = ",M[d_{X2}]]};
Print[" ALICE private auxiliary--------------------
-------"];
{Print[" x1 = ",M[x_1]],Print[" x2 = ",M[x_2]]};
Print[" ALICE private keys ------------------------
-------"];
{Print[" a1 = ",M[a_1]],Print[" a2 = ",M[a_2]],Print["
a3 = ",M[a_3]]};
Print[" ALICE public keys (=token for BOB:{u,v,w}---
-------"];
{Print[" u = ",M[u]],Print[" v = ",M[v]],Print[" w =
",M[w]]};
Print[" Random BOB private eigenvalues ------------
--------"];
{Print[" dB1 = ",M[d_{B1}]],Print[" dB2 = ",M[d_{B2}]]};
{Print[" dY1 = ",M[d_{Y1}]],Print[" dY2 = ",M[d_{Y2}]]};
Print[" BOB private auxiliary----------------------
-------"];
{Print[" Y1 = ",M[y_1]],Print[" Y2 = ",M[y_2]]};
Print[" BOB private keys --------------------------
-------"];
{Print[" b1 = ",M[b_1]],Print[" b2 = ",M[b_2]],Print["
b3 = ",M[b_3]]};
Print[" BOB public keys (=token for ALICE:{p,q,r}---
-------"];
{Print[" p = ",M[p]],Print[" q = ",M[q]],Print[" r =
",M[r]]};
Print[" ALICE & BOB (identical) session
keys/conjugators--"];
{Print[" Kalice = ",M[Kalice]],Print[" Kbob  =
",M[Kbob]]};
Print[" PKE (BCSP) session-------------------------
------"];
Print[" BOB selected msg = "M[msg]];
Print[" Encrypted msg = "M[cif]];
Print[" ALICE recovered msg = "M[rec]];
Print[" -------------------------------------------
------"];
```

**LAST SESSION OUTPUT**

----------------------------------------
**Matrix dimension (dim) = 8**
**Prime modulus (prime)  = 251**
**Number of Sessions     = 1000**
**Mean session time (sec)= 0.001968`**
**Sessions singularity repair needed = 0.1%**
----------------------------------------

**Random public eigenvectors ----------------**
P (for A2,Y1) =
   {{118, 87, 80, 104, 185, 139, 159, 46},
   {154, 244, 21, 132, 150, 119, 37, 202},
   {209, 129, 161, 232, 12, 95, 13, 117},
   {146, 223, 33, 122, 12, 173, 170, 205},
   {27, 229, 5, 186, 217, 136, 13, 9},
   {25, 179, 38, 239, 172, 50, 125, 108},
   {176, 112, 200, 108, 145, 171, 76, 100},
   {2, 45, 164, 12, 201, 73, 154, 199}}
Q (for A3,Y2) =
   {{249, 119, 144, 39, 171, 239, 154, 249},
   {205, 165, 39, 7, 76, 121, 22, 44},
   {235, 42, 74, 189, 25, 123, 58, 42},
   {162, 108, 52, 101, 126, 69, 10, 217},
   {226, 225, 56, 178, 206, 145, 217, 113},
   {10, 191, 141, 66, 5, 30, 94, 91},
   {90, 129, 91, 230, 47, 156, 35, 148},
   {73, 63, 127, 91, 40, 201, 139, 133}}
R (for B1,X1) =
   {{35, 46, 161, 77, 163, 16, 49, 12},
   {244, 187, 177, 46, 169, 186, 198, 164},
   {105, 78, 198, 72, 188, 30, 14, 241},
   {195, 69, 14, 241, 52, 18, 114, 130},
   {186, 129, 205, 26, 229, 232, 207, 205},
   {186, 83, 244, 163, 215, 29, 12, 237},
   {238, 99, 79, 37, 32, 91, 248, 96},
   {13, 169, 7, 195, 25, 228, 50, 127}}
S (for B2,X2) =
   {{39, 8, 145, 170, 38, 159, 197, 118},
   {41, 66, 220, 106, 243, 52, 50, 114},
   {9, 7, 128, 145, 113, 137, 184, 228},
   {214, 126, 35, 170, 49, 38, 249, 230},
   {170, 81, 136, 8, 98, 92, 175, 7},
   {119, 71, 173, 142, 247, 91, 146, 168},
   {191, 214, 36, 44, 82, 211, 199, 2},
   {105, 32, 206, 102, 27, 76, 139, 198}}
**Random ALICE private eigenvalues --------**
dA2 =
   {{39, 0, 0, 0, 0, 0, 0, 0},
   {0, 56, 0, 0, 0, 0, 0, 0},
   {0, 0, 214, 0, 0, 0, 0, 0},
   {0, 0, 0, 120, 0, 0, 0, 0},
   {0, 0, 0, 0, 97, 0, 0, 0},
   {0, 0, 0, 0, 0, 56, 0, 0},
   {0, 0, 0, 0, 0, 0, 110, 0},
   {0, 0, 0, 0, 0, 0, 0, 87}}}
dA3 =
   {{250, 0, 0, 0, 0, 0, 0, 0},
   {0, 214, 0, 0, 0, 0, 0, 0},
   {0, 0, 226, 0, 0, 0, 0, 0},
   {0, 0, 0, 16, 0, 0, 0, 0},
   {0, 0, 0, 0, 26, 0, 0, 0},
   {0, 0, 0, 0, 0, 186, 0, 0},
   {0, 0, 0, 0, 0, 0, 22, 0},
   {0, 0, 0, 0, 0, 0, 0, 248}}
dX1 =
   {{62, 0, 0, 0, 0, 0, 0, 0},
   {0, 46, 0, 0, 0, 0, 0, 0},
   {0, 0, 10, 0, 0, 0, 0, 0},
   {0, 0, 0, 80, 0, 0, 0, 0},
   {0, 0, 0, 0, 88, 0, 0, 0},
   {0, 0, 0, 0, 0, 175, 0, 0},
   {0, 0, 0, 0, 0, 0, 195, 0},
   {0, 0, 0, 0, 0, 0, 0, 28}}

dX2 =
   {{230, 0, 0, 0, 0, 0, 0, 0},
   {0, 209, 0, 0, 0, 0, 0, 0},
   {0, 0, 186, 0, 0, 0, 0, 0},
   {0, 0, 0, 31, 0, 0, 0, 0},
   {0, 0, 0, 0, 236, 0, 0, 0},
   {0, 0, 0, 0, 0, 31, 0, 0},
   {0, 0, 0, 0, 0, 0, 241, 0},
   {0, 0, 0, 0, 0, 0, 0, 117}}

**ALICE private auxiliary-----------------**
x1 =
   {{185, 177, 6, 133, 82, 219, 154, 65},
   {160, 161, 244, 214, 15, 99, 58, 81},
   {63, 24, 105, 6, 116, 131, 213, 28},
   {185, 23, 148, 104, 148, 95, 40, 109},
   {182, 144, 0, 191, 5, 230, 179, 206},
   {216, 235, 144, 109, 170, 177, 234, 159},
   {13, 214, 202, 190, 209, 100, 229, 78},
   {39, 135, 206, 239, 29, 145, 5, 220}}
x2 =
   {130, 56, 107, 168, 26, 102, 76, 27},
   {140, 71, 189, 9, 211, 248, 64, 199},
   {197, 207, 146, 173, 121, 54, 49, 165},
   {159, 139, 163, 40, 76, 232, 15, 242},
   {50, 91, 127, 81, 154, 242, 80, 228},
   {28, 75, 51, 213, 126, 145, 166, 212},
   {146, 26, 53, 16, 11, 32, 24, 32},
   {44, 236, 139, 214, 53, 100, 164, 69}}
**ALICE private keys ---------------------**
a1 =
   {82, 187, 30, 241, 138, 17, 139, 234},
   {138, 221, 8, 88, 14, 226, 200, 110},
   {5, 151, 124, 207, 119, 38, 86, 96},
   {5, 248, 171, 240, 43, 10, 147, 164},
   {100, 123, 237, 224, 53, 57, 185, 143},
   {107, 13, 45, 90, 45, 191, 17, 188},
   {42, 147, 60, 95, 75, 157, 235, 157},
   {82, 58, 89, 222, 86, 205, 20, 38}}
a2 =
   {{221, 163, 237, 236, 25, 117, 236, 30},
   {53, 50, 223, 72, 124, 31, 45, 136},
   {151, 82, 154, 14, 120, 110, 69, 16},
   {115, 65, 61, 114, 16, 26, 203, 70},
   {236, 93, 237, 33, 175, 113, 237, 46},
   {185, 192, 24, 203, 58, 132, 59, 8},
   {169, 199, 85, 126, 229, 210, 45, 191},
   {141, 134, 123, 0, 229, 227, 72, 139}}
a3 =
   {{63, 41, 113, 40, 99, 102, 22, 156},
   {1, 128, 57, 143, 48, 49, 157, 100},
   {43, 68, 23, 245, 109, 22, 158, 217},
   {5, 223, 161, 138, 0, 67, 218, 33},
   {113, 211, 77, 96, 165, 181, 209, 148},
   {136, 156, 168, 192, 100, 60, 205, 88},
   {81, 174, 184, 104, 209, 164, 16, 25},
   {74, 68, 154, 209, 131, 184, 225, 93}}

**ALICE public keys (=token for BOB:{u,v,w}---**
u =
   {{13, 45, 16, 222, 244, 44, 32, 228},
   {137, 52, 66, 33, 117, 121, 223, 96},
   {249, 3, 160, 173, 208, 137, 211, 89},
   {117, 129, 223, 30, 245, 6, 26, 170},
   {246, 4, 33, 156, 215, 79, 41, 185},
   {218, 149, 161, 54, 184, 85, 149, 250},
   {200, 43, 17, 4, 225, 165, 44, 228},
   {22, 161, 92, 201, 223, 46, 17, 63}}
v =
   {{151, 241, 216, 80, 197, 30, 29, 56},
   {57, 32, 151, 200, 39, 9, 196, 90},
   {0, 16, 104, 112, 88, 184, 128, 126},
   {32, 167, 46, 6, 211, 248, 67, 61},
   {2, 136, 41, 207, 181, 241, 210, 112},
   {157, 167, 14, 226, 34, 131, 13, 202},
   {165, 225, 89, 144, 70, 143, 104, 199},
   {43, 54, 47, 76, 58, 76, 218, 217}}

```
w =
   {{112, 5, 86, 148, 45, 217, 142, 68},
    {0, 203, 179, 25, 32, 185, 160, 98},
    {10, 97, 83, 114, 6, 145, 118, 156},
    {96, 41, 30, 93, 184, 109, 115, 59},
    {0, 224, 44, 104, 128, 191, 142, 91},
    {157, 49, 41, 196, 73, 58, 191, 72},
    {200, 231, 240, 0, 96, 209, 45, 124},
    {22, 4, 201, 127, 151, 31, 108, 68}}
Random BOB private eigenvalues ----------
dB1 =
   {{125, 0, 0, 0, 0, 0, 0, 0},
    {0, 103, 0, 0, 0, 0, 0, 0},
    {0, 0, 197, 0, 0, 0, 0, 0},
    {0, 0, 0, 99, 0, 0, 0, 0},
    {0, 0, 0, 0, 29, 0, 0, 0},
    {0, 0, 0, 0, 0, 178, 0, 0},
    {0, 0, 0, 0, 0, 0, 45, 0},
    {0, 0, 0, 0, 0, 0, 0, 56}}
dB2 =
   {{117, 0, 0, 0, 0, 0, 0, 0},
    {0, 6, 0, 0, 0, 0, 0, 0},
    {0, 0, 191, 0, 0, 0, 0, 0},
    {0, 0, 0, 82, 0, 0, 0, 0},
    {0, 0, 0, 0, 148, 0, 0, 0},
    {0, 0, 0, 0, 0, 78, 0, 0},
    {0, 0, 0, 0, 0, 0, 244, 0},
    {0, 0, 0, 0, 0, 0, 0, 247}}
dY1 =
   {{215, 0, 0, 0, 0, 0, 0, 0},
    {0, 142, 0, 0, 0, 0, 0, 0},
    {0, 0, 12, 0, 0, 0, 0, 0},
    {0, 0, 0, 123, 0, 0, 0, 0},
    {0, 0, 0, 0, 241, 0, 0, 0},
    {0, 0, 0, 0, 0, 234, 0, 0},
    {0, 0, 0, 0, 0, 0, 84, 0},
    {0, 0, 0, 0, 0, 0, 0, 13}}
dY2 =
   {{11, 0, 0, 0, 0, 0, 0, 0},
    {0, 77, 0, 0, 0, 0, 0, 0},
    {0, 0, 95, 0, 0, 0, 0, 0},
    {0, 0, 0, 145, 0, 0, 0, 0},
    {0, 0, 0, 0, 172, 0, 0, 0},
    {0, 0, 0, 0, 0, 162, 0, 0},
    {0, 0, 0, 0, 0, 0, 235, 0},
    {0, 0, 0, 0, 0, 0, 0, 40}}
BOB private auxiliary------------------
Y1 =
   {{103, 68, 213, 144, 111, 124, 138, 81},
    {195, 112, 28, 130, 186, 165, 11, 108},
    {57, 207, 123, 113, 41, 250, 110, 162},
    {154, 5, 58, 163, 104, 127, 80, 14},
    {80, 238, 220, 173, 122, 216, 155, 159},
    {88, 23, 149, 110, 197, 226, 119, 202},
    {135, 109, 47, 28, 23, 91, 134, 154},
    {163, 19, 71, 107, 235, 187, 187, 81}}
Y2 =
   {{143, 44, 16, 226, 219, 211, 247, 34},
    {119, 246, 212, 130, 139, 29, 128, 226},
    {115, 231, 80, 30, 164, 204, 17, 167},
    {6, 5, 7, 169, 80, 246, 232, 246},
    {7, 240, 220, 162, 32, 93, 213, 9},
    {176, 146, 74, 187, 141, 229, 155, 181},
    {152, 108, 10, 77, 11, 249, 195, 227},
    {237, 145, 67, 33, 124, 130, 14, 94}}
BOB private keys ------------------------
b1 =
   {{110, 36, 153, 231, 129, 169, 61, 132},
    {169, 199, 2, 96, 180, 10, 242, 194},
    {113, 138, 30, 161, 50, 42, 16, 48},
    {229, 80, 105, 44, 10, 146, 213, 63},
    {80, 168, 216, 230, 36, 151, 180, 94},
    {166, 2, 111, 250, 81, 182, 46, 13},
    {45, 194, 235, 16, 202, 126, 106, 31},
    {106, 110, 205, 91, 190, 54, 218, 125}}
b2 =
   {{115, 14, 121, 94, 8, 143, 86, 207},
    {15, 144, 96, 91, 111, 24, 190, 201},
    {83, 218, 53, 165, 205, 60, 223, 72},
    {42, 46, 72, 4, 69, 79, 56, 10},
    {68, 15, 59, 113, 96, 48, 137, 114},
    {65, 10, 51, 38, 14, 172, 190, 4},
    {20, 144, 218, 18, 27, 20, 9, 52},
    {175, 226, 234, 20, 245, 32, 201, 18}}
b3 =
   {{52, 14, 114, 160, 182, 49, 74, 113},
    {228, 167, 208, 164, 46, 86, 205, 219},
    {18, 51, 40, 202, 18, 174, 96, 191},
    {38, 17, 40, 22, 188, 65, 174, 185},
    {47, 70, 17, 49, 12, 4, 134, 236},
    {57, 180, 82, 103, 250, 29, 148, 148},
    {176, 32, 31, 7, 97, 66, 79, 39},
    {67, 161, 100, 125, 24, 237, 218, 42}}
BOB public keys (=token for ALICE:{p,q,r}---
p =
   {{120, 234, 218, 175, 137, 230, 107, 93},
    {82, 129, 169, 103, 114, 12, 185, 242},
    {154, 246, 131, 120, 130, 73, 183, 243},
    {204, 84, 4, 38, 223, 198, 124, 99},
    {197, 153, 28, 164, 243, 177, 66, 111},
    {19, 208, 146, 215, 53, 79, 58, 54},
    {17, 86, 38, 243, 123, 183, 124, 242},
    {65, 230, 120, 100, 81, 49, 97, 209}}

q =
   {{43, 149, 88, 213, 211, 66, 168, 142},
    {56, 47, 70, 201, 94, 154, 65, 138},
    {134, 151, 195, 64, 108, 99, 199, 208},
    {0, 239, 107, 125, 206, 213, 28, 90},
    {90, 44, 202, 98, 184, 24, 78, 108},
    {64, 199, 60, 218, 248, 109, 84, 44},
    {85, 106, 117, 36, 50, 238, 166, 66},
    {202, 187, 161, 50, 41, 246, 242, 6}
r =
   {{237, 165, 140, 101, 132, 233, 82, 227},
    {25, 92, 181, 209, 180, 26, 41, 235},
    {32, 24, 123, 58, 178, 237, 136, 190},
    {47, 39, 123, 86, 54, 69, 197, 168},
    {57, 144, 210, 177, 123, 19, 74, 153},
    {2, 111, 18, 81, 185, 218, 191, 128},
    {159, 100, 142, 53, 13, 154, 69, 171},
    {233, 88, 194, 241, 143, 98, 54, 53}}
ALICE & BOB (identical) keys or conjugators--
Kalice =
   {{142, 192, 38, 42, 56, 123, 248, 215},
    {86, 89, 216, 109, 223, 54, 66, 135},
    {88, 206, 63, 134, 249, 39, 87, 2},
    {217, 202, 79, 240, 131, 61, 13, 213},
    {62, 67, 72, 46, 219, 51, 113, 100},
    {17, 234, 189, 210, 242, 230, 86, 193},
    {246, 157, 234, 27, 124, 138, 23, 127},
    {131, 35, 240, 116, 190, 144, 174, 90}}
Kbob   =
   {{142, 192, 38, 42, 56, 123, 248, 215},
    {86, 89, 216, 109, 223, 54, 66, 135},
    {88, 206, 63, 134, 249, 39, 87, 2},
    {217, 202, 79, 240, 131, 61, 13, 213},
    {62, 67, 72, 46, 219, 51, 113, 100},
    {17, 234, 189, 210, 242, 230, 86, 193},
    {246, 157, 234, 27, 124, 138, 23, 127},
    {131, 35, 240, 116, 190, 144, 174, 90}}
-----------------------------------------
UNTIL HERE IT WORKS AS A D-H KE.
FOLLOWING IS A BCSP BASED ENCRYPTION.
-----------------------------------------
```

```
PKE (BCSP) session-------------------------
BOB selected msg =
   {{38, 50, 241, 209, 242, 186, 128, 113},
    {200, 43, 145, 57, 52, 145, 76, 229},
    {78, 58, 70, 144, 45, 161, 100, 101},
    {223, 117, 213, 2, 184, 236, 91, 245},
    {136, 160, 210, 11, 197, 44, 239, 54},
    {233, 226, 126, 139, 7, 246, 165, 48},
    {140, 135, 172, 34, 37, 183, 21, 202},
    {176, 130, 203, 141, 49, 0, 161, 5}}
Encrypted msg =
   {{7, 41, 3, 224, 146, 175, 243, 114},
    {168, 22, 11, 103, 83, 91, 24, 179},
    {113, 16, 19, 249, 128, 231, 87, 176},
    {122, 183, 20, 2, 219, 96, 229, 144},
    {46, 30, 198, 139, 4, 240, 27, 56},
    {146, 5, 221, 58, 234, 184, 77, 191},
    {212, 241, 48, 5, 23, 40, 150, 21},
    {144, 12, 79, 177, 154, 45, 115, 234}}
ALICE recovered msg =
   {{38, 50, 241, 209, 242, 186, 128, 113},
    {200, 43, 145, 57, 52, 145, 76, 229},
    {78, 58, 70, 144, 45, 161, 100, 101},
    {223, 117, 213, 2, 184, 236, 91, 245},
    {136, 160, 210, 11, 197, 44, 239, 54},
    {233, 226, 126, 139, 7, 246, 165, 48},
    {140, 135, 172, 34, 37, 183, 21, 202},
    {176, 130, 203, 141, 49, 0, 161, 5}}
----------------------------------------
```